\documentclass[fleqn]{aa}  

\usepackage{graphicx}
\usepackage{xcolor}
\usepackage{hyperref}
\usepackage{txfonts}
\usepackage{calrsfs}
\usepackage[T1]{fontenc}
\usepackage[mathcal]{eucal}
\usepackage{longtable}
\newcommand{\mum}{\mu_{\rm m}}
\newcommand{\Msun}{\,\hbox{$\rm M_{\odot}$}}
\newcommand{\Lsun}{\, {\rm L}_{\odot}}
\newcommand{\Rsun}{\,\hbox{$\rm R_{\odot}$}}
\newcommand{\Msunpc}{{\rm M}_{\odot}\, {\rm pc}}

\newcommand{\micron}{\hbox{$\mu$m}}

\begin{document}


    \title{Extending the cosmic distance ladder two orders of magnitude with strongly lensed Cepheids, carbon AGB, and RGB stars.}
  
  \titlerunning{High-z SCs}
  \authorrunning{J.M. Diego}

   \author{J. M. Diego \inst{1}
   \and S. P. Willner \inst{2}
   \and J. M. Palencia \inst{1}
   \and R. A. Windhorst \inst{3} 
   }      
  \institute{Instituto de F\'isica de Cantabria (CSIC-UC). Avda. Los Castros s/n. 39005 Santander, Spain 
  \and Center for Astrophysics \textbar\ Harvard \& Smithsonian, 60 Garden St., Cambridge, MA 02138 USA  
  \and School of Earth and Space Exploration, Arizona State University, Tempe, AZ 85287-6004, USA
  }

 \abstract{Gravitational lensing by galaxy clusters can create extreme magnification near the cluster caustics, thereby enabling  detection of individual luminous stars in high-redshift background galaxies. Those stars can include non-explosive standard candles such as Cepheid variables, carbon stars in the asymptotic giant branch, and stars at the tip of the red-giant branch out to $z\lesssim1$. A large number of such detections, combined with modeling of the magnification affecting these stars (including microlensing), opens the door to extending the distance range of these standard candles by two orders of magnitude, thereby providing a check on the distances derived from supernovae.  Practical measurement of a distance modulus depends on measuring the apparent magnitude of a ``knee feature'' in the lensed luminosity function due to the great abundance of red-giant-branch stars just below the luminosity of the tip of the red-giant branch. 
 As a bonus, strongly lensed stars detected in deep exposures also provide a robust method of mapping small dark-matter substructures, detections of which also cluster around the critical curves of small-scale dark matter halos.  
   }

   \maketitle
%

\section{Introduction}
A puzzling tension in cosmology is the apparent discrepancy in the observed expansion rate of the universe, as measured by the Hubble constant $H_0$,  when this rate is measured through different observations. Fluctuations in the cosmic microwave background combined with the spatial correlation between galaxies when the universe was half its present age prefer a relative slow rate of expansion $H_0\approx 68$ km\, s$^{-1}$\, Mpc$^{-1}$ \citep{Planck2020}. On the other hand, measurements of distances to distant supernovae (SNe), anchored by distance measurements to nearby standard candles such as Cepheid stars,  prefer a faster rate of expansion $H_0\approx 74$ km\, s$^{-1}$\, Mpc$^{-1}$ \citep{Riess2019}. Considering the uncertainty in both measurements, the tension between the two stands currently at ${\gtrsim} 4\sigma$. \\

Multiple efforts to make more precise measurements of $H_0$ using alternative methods are underway, often involving standard candles (SCs) other than SNe. Other SCs include Cepheids, asymptotic-giant-branch (AGB) carbon stars, and the tip of the red-giant branch (TRGB)\null. These are used to estimate distances to nearby galaxies and thus calibrate the SN distance ladder \citep{Riess2019,Freedman2024}. Here we refer to these SCs as ``non-explosive SCs or NESCs''  to differentiate them from SNe. 
The range of distances over which  NESCs have been applied is only a few tens of Mpc, limited by their luminosity, which is orders of magnitude below that of type Ia SNe.  \\

In spite of NESCs' low luminosities, in specific circumstances NESCs can be observed at cosmological distances comparable to those where most SNe are found. Thanks to extreme magnification ($\mu>1000$) near the critical curves (CC) of galaxy clusters, luminous stars can be detected even at high redshift. The first high-$z$ star discovered this way was Icarus at $z=1.49$ \citep{Kelly2018}, which was quickly followed by many more examples \citep{Chen2019,Kaurov2019,DiegoGodzilla,Diego2023Gordo,Furtak2024} extending the range of redshifts of extremely magnified stars up to $z\approx 6$ with Earendel \cite{Welch2022}. These stars have lensing-corrected luminosities $10^4 \lesssim \Lsun \lesssim 10^5$, comparable to NESC luminosities, but so far no high-$z$ NESC has been recognized. One reason is that this requires light curves with multiple epochs for pulsating stars such as Cepheids or finding many stars in the same galaxy to identify subgroups in color--magnitude diagrams (CMD)\null. A second reason is that so far few observations reach the needed depth: most NESCs are expected to be very faint, even for JWST, despite the large magnification factors present near the CC of galaxy clusters.\\ 

Despite the difficulties, recent observations by JWST are starting to show large numbers of luminous stars in individual galaxies at $z>0.5$. The most spectacular example so far is the Dragon arc, a $z=0.725$ galaxy amplified by the $z=0.375$ galaxy cluster A370. \cite{Fudamoto2024} identified $\approx$45 transients in difference images of two epochs with a 5$\sigma$ detection limit of $\approx$28.75 AB in F200W (observed wavelength 2.0~\micron) difference images. Most of these Dragon transients are believed to be microlensing events of very luminous stars ($M_{J}<-6.5$).  (Observed 2.0~\micron\ corresponds to rest 1.16~\micron\ or approximately $J$-band.) Detecting these stars is possible only because a sizable portion of the Dragon galaxy is being magnified by extreme values. Macromodel magnification $\mum >100$ covers an  area ${\ga} 190$~kpc$^2$ in the lens plane overlapping the Dragon.  This area is large enough to make microlensing events common \citep{Diego2024a}. During a microlensing event, the magnification can exceed a few thousand for a few days, making the brightest stars in the Dragon detectable during that period. In the source plane, and accounting for the multiplicity of the lensed images, the area with $\mum >100$ is $\approx$0.5~kpc$^2$. As a comparison, the Large Magellanic Cloud contains $\approx$200 bright AGB stars per kpc$^2$ \citep{Cioni2006}. The near-infrared (NIR) stellar population of AGB stars forms a series of branches \citep{Weinberg2001}, which contain some of the most popular SCs (Figure~1). Bright stars in the 0.5~kpc$^2$ of high magnification in the Dragon have a high probability of experiencing microlensing events by regular stars in the intracluster medium (ICM) of A370, and these will appear as transient events in multi-epoch observations. Among the brightest stars ($M_{J}<-4$) in the Dragon, deep observations will reveal a wealth of NESCs, including long-period Cepheids, carbon AGB stars, and TRGB stars. 
TRGBs constitute the faintest group in this sample at an estimated magnitude ($M_{J}^{\rm TRGB}<-4.1$ \citep{Newman2024,Anand2024}. AGB stars in the J-Region (or JAGB stars) are expected to be $\approx$1~mag brighter \citep{Madore2020,Ripoche2020,Freedman2024}.  
Finally, long-period Cepheids can be a few magnitudes brighter still. Their absolute magnitude in the $J$-band relates to their period as: $M_J = -3.18[{\log}_{10}(P)-1.0] -5.22$, where $P$ is expressed in days \citep{Storm2011}. 
The longest-period  Cepheids have $P> 50$~days and can have $M_J < -7.4$. Cepheids with period $\approx 5$ days are comparable in luminosity to TRGB stars.
During a microlensing event, the magnification can exceed 5000, making a TRGB star at $z=0.725$ with absolute magnitude $M_J=-4.1$ (AB) acquire an apparent magnitude 29.85 in JWST's F200W filter, detectable in exposures of $\approx$4~hours. Carbon AGB stars and long-period Cepheids can also be detected in these deep observations of the Dragon, opening the door to an independent estimate of the distance to the Dragon based on NESCs.  
All these stars can momentarily be microlensed and detected as transients in deep JWST multiepoch observations.  Some of these stars, such as the Cepheids, can be  detected as transients even in the absence of microlensing because of their intrinsic varying nature.  \\

This paper investigates the feasibility of detecting NESC stars in galaxies at $z>0.5$ and in particular in the Dragon arc. We pay special attention to the dilution of the markers used in local measurements (as in the JAGB and TRGB methods) due to the wide range of magnification factors affecting the observed lensed luminosity function (LF) and CMDs. 
The paper is organized as follows.
Sect.~\ref{Sec_LMC} presents the sample of stars used to study the effects of magnification at high redshift. 
The probability of magnification is briefly discussed in Sect.~\ref{Sec_Mu}. 
Section~\ref{Sec_CMD} shows the resulting CMD of a population of bright stars at $z=0.725$ that is being magnified by the galaxy cluster and microlensed by stars in the ICM\null. 
Section~\ref{Sec_LF} presents the expected  of the lensed stars in the Dragon arc. Section~\ref{Sec_Discussion} discusses some implications , and Sect.~\ref{Sec_Conclusions} gives a summary. 
A flat cosmological model was adopted with $\Omega_{\rm M} =0.3$ but with two values for the Hubble constant, $H_0=68$~km~s$^{-1}$ Mpc$^{-1}$ (or $h=0.68$) and $H_0=74$~m~s$^{-1}$ Mpc$^{-1}$ (or $h=0.74$). For these cosmological models, the distance moduli for a galaxy at $z=0.725$ are 43.30 and 43.12, respectively. The `macromodel magnification'  $\mum$ is the magnification due to the galaxy cluster without accounting for the distortion in the magnification due to microlenses. These microlenses produce very narrow regions of divergent magnification in the source plane known as `microcaustics'.

\section{Bright stars in the LMC from 2MASS}\label{Sec_LMC}

\begin{figure} 
   \includegraphics[width=9.0cm]{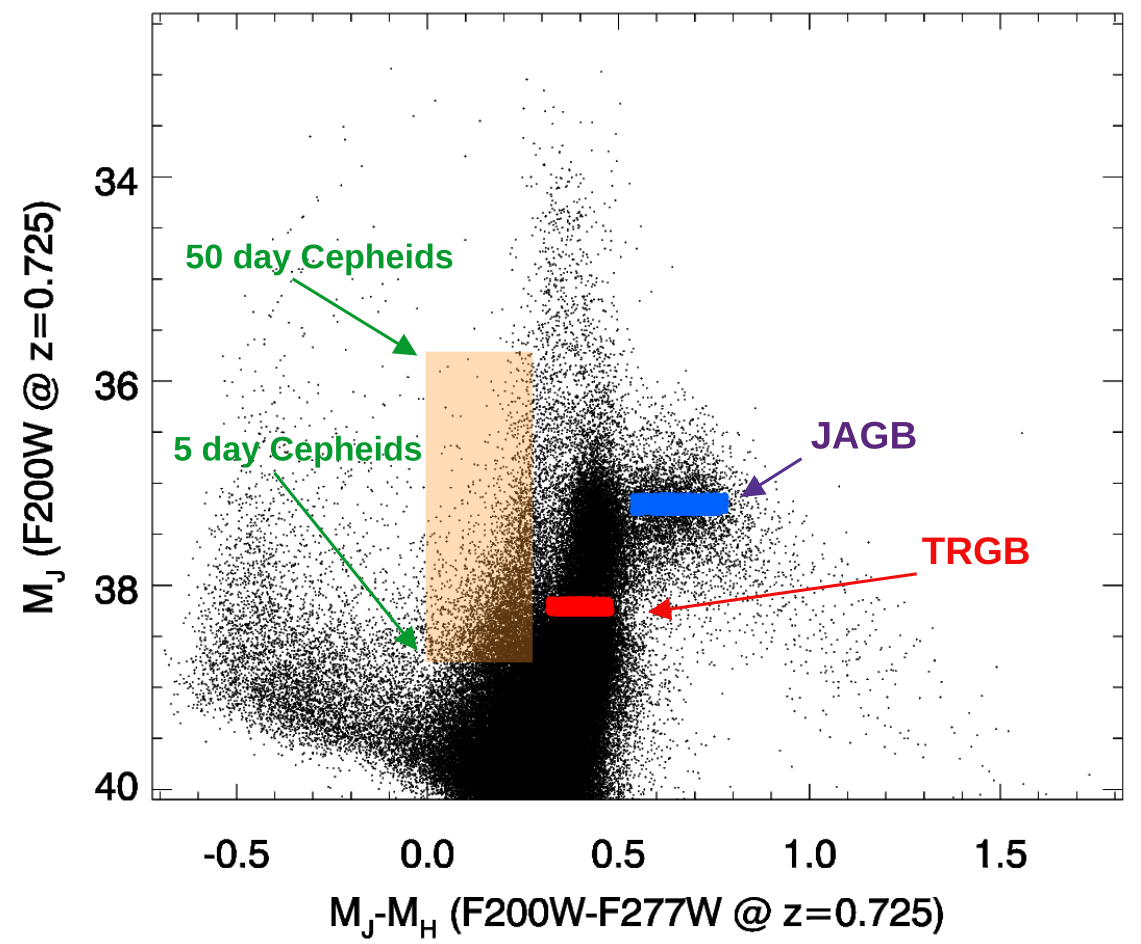}  
      \caption{CMD of stars in a small portion of the LMC as observed by 2MASS in the $J$ and $H$ bands. The magnitudes have been corrected by distance assuming the stars are at $z=0.725$ but not magnified. Popular SCs are marked:  the J-region in blue, the TRGB in red, and Cepheids in the orange rectangle. The TRGB appears at $M_{ J}\approx 39$. The number of stars brighter than apparent magnitude 40 is $N_{*}\approx 120000$.}
         \label{Fig_LMC_AGB}
\end{figure}

To simulate a population of bright stars at $z=0.725$, we consider real observations of stars in the Large Magellanic Cloud (LMC) observed with 2MASS\null  \citep{Ripoche2020}. At $z=0.725$, the $J$ band redshifts into JWST's F200W filter while the $H$ band redshifts into the F277W filter. These two filters are the most sensitive to cool stars at this redshift. We transformed the 2MASS Vega magnitudes to AB magnitudes following \cite{Maiz2007} (their Table 4) and place the stars at $z=0.725$ by correcting for the distance modulus. The resulting CMD is shown in Figure~\ref{Fig_LMC_AGB}. The different branches identified in the $J$ and $K$ bands \cite{Weinberg2001} are also clearly detected in the $J$ and $H$ bands because the $H-K$ color has a relatively small range. In particular the carbon AGB stars form a well defined compact group. At $z=0.725$, the TRGB  appears at  $M_{J}^{TRGB}\approx 39$ for $h=0.74$ and 0.18 magnitudes fainter for $h=0.68$.  
At similar apparent magnitude we expect to see classic Cepheids with periods $P\approx5$ days. Longer period Cepheids with $P=50$ days are expected to appear $\approx 3$ magnitudes brighter.  Brighter Cepheids with even longer periods  exist but are exceedingly rare and  unlikely to be found in the  area of large magnification. 
After being magnified by the cluster and microlenses, the magnitudes of all stars spread upwards, but their colors remain unchanged  as described in more detail in Sect.~\ref{Sec_CMD} below. 

\section{Probability of Magnification}\label{Sec_Mu}
The magnification experienced by  luminous stars in the Dragon is a combination of the macromodel magnification and magnification by microlenses in the A370 cluster's ICM\null. For microlensing to take place, a microlens' distortion in the magnification pattern must intersect the line of sight to a star in the Dragon. The probability that this will happen can be thought of as an optical depth, where  ``optically thick'' means the probability is high.
The transition to optically thick takes place when the effective surface-mass-density of microlenses is close to the critical surface-mass-density, $\Sigma_{\rm eff}\equiv\mum \Sigma_{*} \approx \Sigma_{\rm crit}$, where $\mum $ is the macromodel magnification and $\Sigma_{*}$ is the microlens surface mass density. 
At the position of the Dragon, $\Sigma_{*} \approx 30\, \Msunpc^{-2}$, and for the redshifts of A370 and the Dragon 
$\Sigma_{\rm crit}= 3640\, \Msunpc^{-2}$ \citep{Diego2024b}. 
This means the optically thick regime corresponds to  $\mum \gtrsim 120$. The macromodel magnification varies along the Dragon arc, weakly along the CC but with strong dependence on $d$, the separation from the CC\null. Typical values of $\mum $ are a few tens when $d$ is a few arcseconds to  $\mum \gg100$ when $d$ is a fraction of an arcsecond. The exact location of the CC is unknown, and different models predict different locations, but most models predict that the magnification scales as $\mum  = A/d$, where $A$ is a parameter that varies along the CC but takes values $30\arcsec \lesssim A \lesssim 100\arcsec$.

\begin{figure} 
   \includegraphics[width=9.0cm]{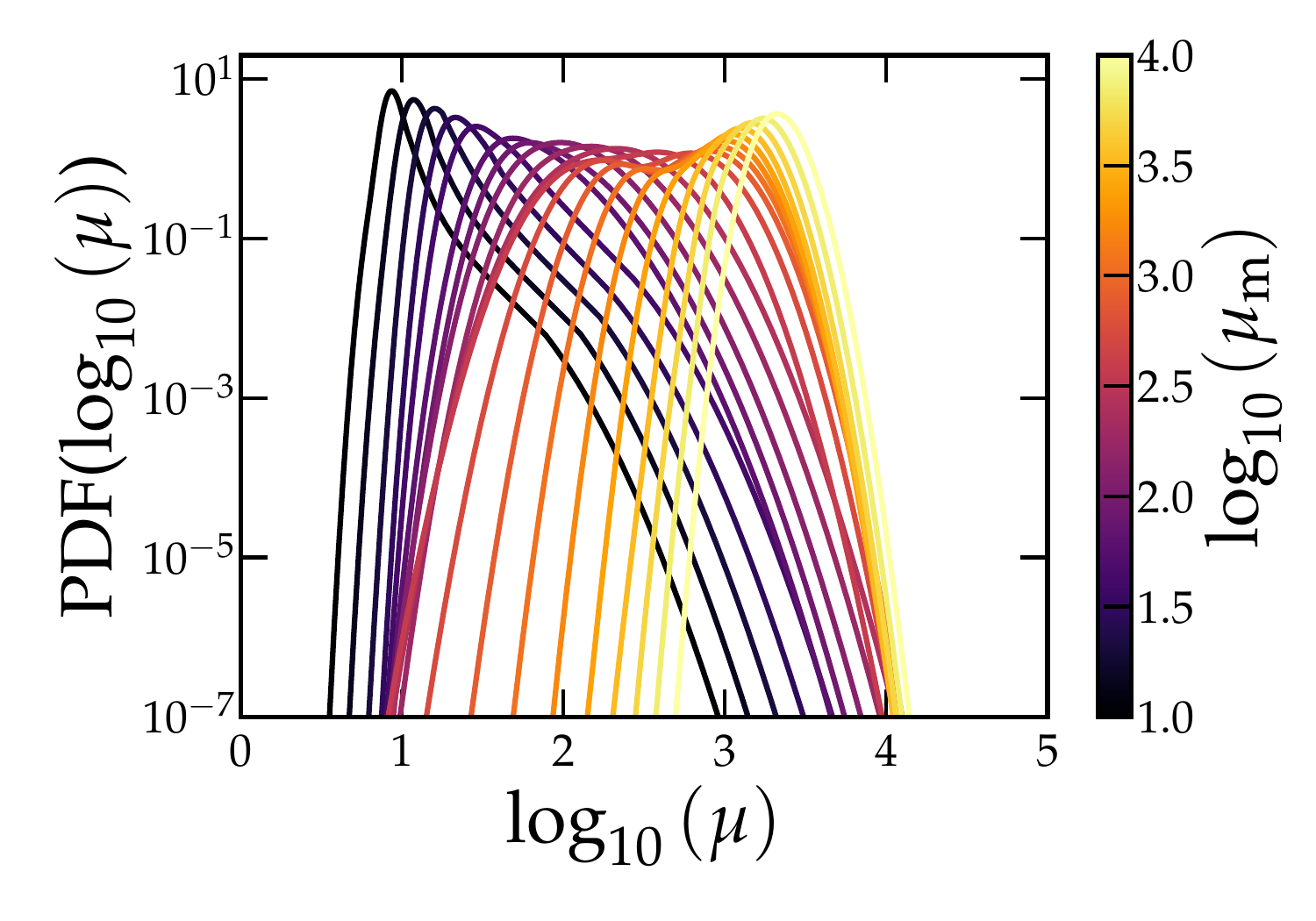}  
      \caption{Probability distribution of total magnification, 
      $P(\mu)$, after accounting for macromodel and microlenses.
      Each curve is for a fixed macromagnification $\mum$ as indicated by the curve's color and the color bar on the right.  As explained in the text, $\mum$ mainly depends on separation from the cluster CC\null. The curves were derived with the code M\_SMiLe \cite{Palencia2023,Palencia2024} which computes the PDF of magnification for any combination of redshifts, $\mum$, and  $\Sigma_{*}$.
      }
         \label{Fig_PDFmu}
\end{figure}

Depending on the value of $\mum$, which mostly depends on the separation $d$ from the cluster CC, the probability distribution function (PDF) of magnification and the role of microlenses varies. At   $d>2\arcsec$, in the optically thin regime,  $\mum  < 100$, and the PDF of the magnification has a relatively narrow peak near $\mum$ and a long tail scaling as $\mu^{-3}$ for magnifications  ${\gg}\mum$. Very close to the CC, when $\mum\gg100$ in the optically thick regime, the microlenses play a very important role in shaping the magnification PDF\null. In this regime, microlenses introduce constant small-scale perturbations in the deflection field and increase its curvature. Because the magnification is inversely proportional to the curvature of the deflection field, the most likely value for the magnification drops below $\mum$, and the PDF of the magnification becomes a lognormal. Between these two extremes, the PDF  transitions between the two limiting shapes as shown in Figure~\ref{Fig_PDFmu}.
One important property is that when $\Sigma_{\rm eff} \gg \Sigma_{\rm crit}$, the PDF becomes a  lognormal similar to the last curve (yellow) in Figure~\ref{Fig_PDFmu}. This  property  can be exploited because  it guarantees that sufficiently close to the CC, all events are affected by the same magnification PDF\null. This enables robust forward modeling of the  of lensed stars through the known magnification PDF to give a direct confrontation with observations. Comparing models for different s allows the characteristics of the NESC population to be extracted. 

The curves shown in Figure~\ref{Fig_PDFmu} were obtained from inverse ray tracing of large simulations but with a finite pixel size comparable to the size of a large star with radius $\approx$300\Rsun. Smaller stars can attain magnifications $\ga$10\,000 when they cross a microcaustic, but that is not reproduced by the curves in Figure~\ref{Fig_PDFmu}. Such enormous magnification factors can typically be maintained for only hours to a few days, depending on the star radius and its velocity relative to the microcaustic. 

\section{The CMD of magnified bright stars}\label{Sec_CMD}
\begin{figure} 
   \includegraphics[width=9.0cm]{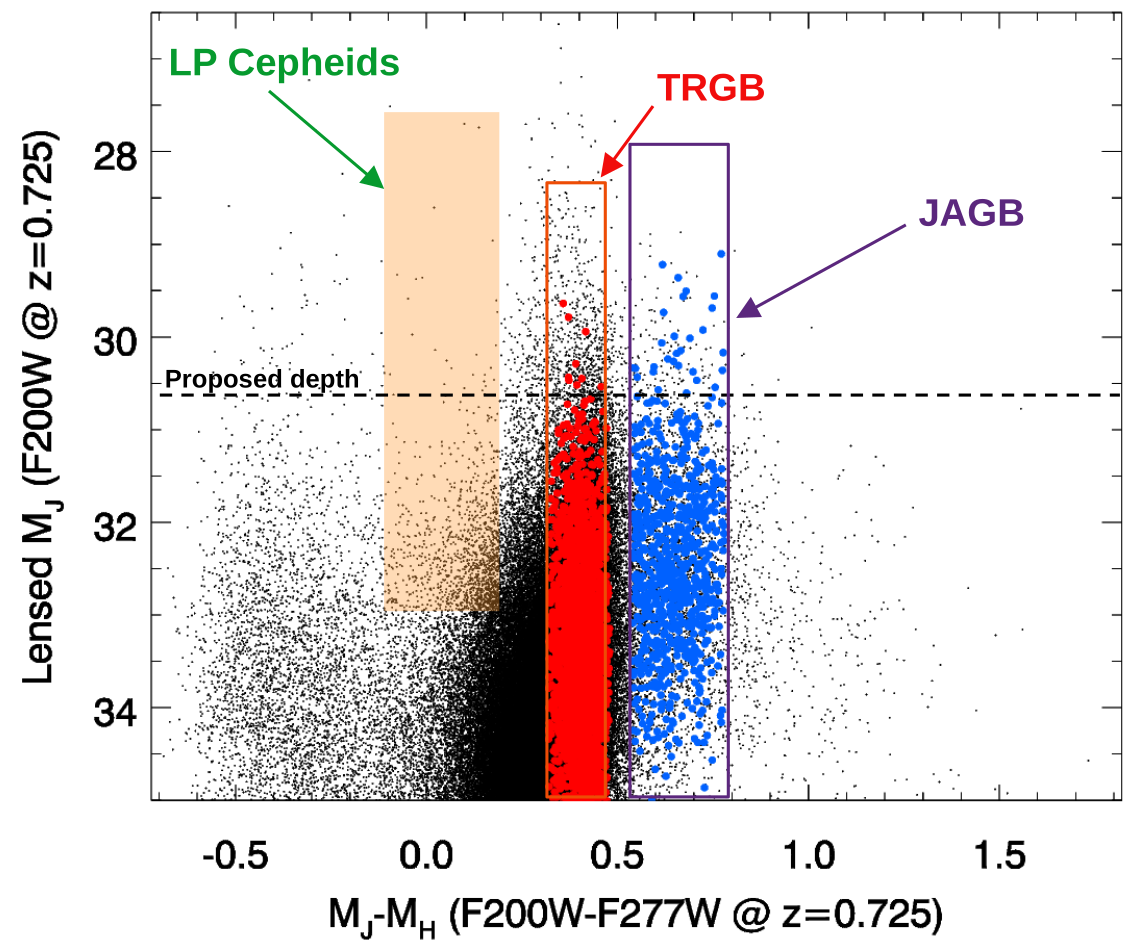}  
      \caption{CMD of the stars in Figure~\ref{Fig_LMC_AGB} at $z=0.725$ after being magnified by A370 and microlenses (Equation~1). The red points correspond to stars in the narrow TRGB magnitude range marked in red in Figure~\ref{Fig_LMC_AGB}, and the blue dots correspond to stars in the blue JAGB region. The random magnifications spread the stars' magnitudes over a  range shown, but the magnification is wavelength-independent.
 The detection limit achievable with JWST in exposures of 16.7 hours in F200W (AB 30.6 mag) is shown as a horizontal dashed line.  The number of stars brighter than apparent magnitude 30.6 $N_{*}\approx 1100$.}
         \label{Fig_LMC_AGB_Lensed}
\end{figure}

To simulate deep JWST observations of stars in an LMC-like galaxy at $z=0.725$  crossing the cluster caustic of A370, we combined the stellar population shown in Figure~\ref{Fig_LMC_AGB} with the magnification PDF shown in Figure~\ref{Fig_PDFmu}. We assigned each star a random macromodel magnification in the range $50<\mum<5000$ following the canonical PDF of macromodel magnification, $P(\mum)\propto\mum^{-3}$ \citep{Schneider1992}. The lower limit $\mum=50$ corresponds to the lowest macromodel magnification expected in portions of the Dragon galaxy where individual stars can still be detected (through microlensing). The upper limit $\mum=5000$ corresponds to the highest peak in the microlensed lognormal pdf shown in Figure~\ref{Fig_PDFmu} expected near the cluster CC\null. This peak is lower than the macromodel magnification near the CC owing to the effect of microlenses  \citep{Palencia2023}. Next, for each $\mum$ we computed $P(\mu_{\rm micro})$ according to the value of $\mum$. Each $P(\mu_{\rm micro})$ resembles a curve in  Figure~\ref{Fig_PDFmu}. Finally, we randomly assigned  $\mu_{\rm micro}$ from the generated $P(\mu_{\rm micro})$ and computed the apparent magnitude after lensing amplification according to 
\begin{equation}
M_{\rm Dragon} = M_{\rm LMC} - (18.5-43.12) -0.59 -2.5*\log_{10}(\mu)
\end{equation}
where $M_{\rm LMC}$ is the observed AB magnitude in $J$ or $H$  \citep{Ripoche2020}, while 18.5 and 43.12 are the distance moduli to the LMC and $z=0.725$ respectively (for h=0.74), and $0.59=2.5{\rm log}_{10}(1+z)$ is the bandwidth correction after making the simplification that at $z=0.725$ the J-band redshifts identically into F200W and that the filter response is similar in both bands.
%
The resulting apparent magnitudes are shown in Figure~\ref{Fig_LMC_AGB_Lensed}. 
For $h=0.68$, all stars would be $\approx$0.18~fainter, but even in that case
many NESCs  from the three groups, Cepheids, JAGB, and TRGB should be above the detection limit.

As discussed at the end of Section~\ref{Sec_Mu}, the magnifications adopted in this calculation assume large stars with $R_{*}=300$\Rsun. Smaller stars, such as those in the TRGB with maximum radius $R_{*}\approx 100$\Rsun, can have maximum magnifications a factor ${(300/R_{*})}^{0.5}$  larger during a microcaustic crossing (that is, 0.75~mag brighter for stars with $R_{*}=75$\Rsun). This will promote a few more NESCs---ones very close to a microcaustic when observed---above the detection limit and slightly increase the numbers shown in Figure~\ref{Fig_LMC_AGB_Lensed}

\section{Lensed luminosity function}\label{Sec_LF}
\begin{figure} 
  \includegraphics[width=9.0cm]{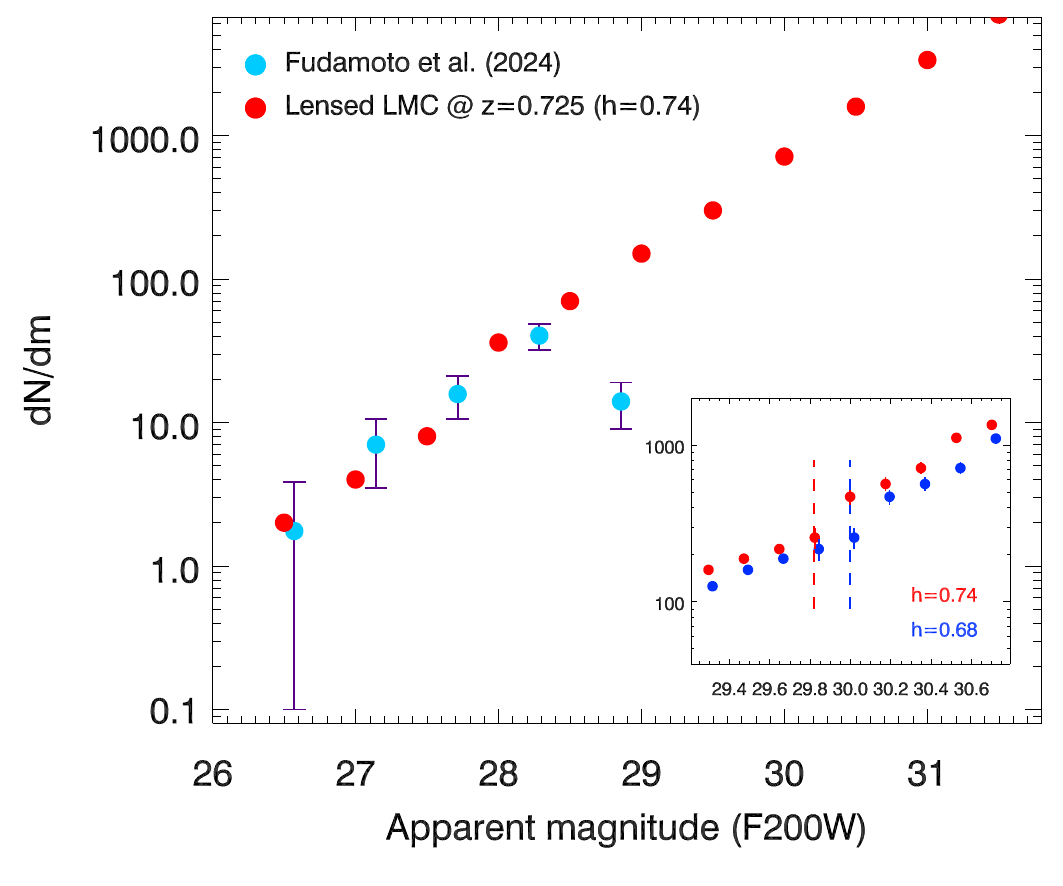}  
      \caption{Luminosity function (LF) of observed transients in the Dragon galaxy \citep{Fudamoto2024} (blue dots) compared with the expected LF (red dots). Error bars on the blue points are from Poisson statistics.  The red points correspond to a randomly selected subsample of 180000 luminous stars from the LMC. The expected LF is that of an LMC-like galaxy at $z=0.725$ in a cosmology with $h=0.74$ (Figure~\ref{Fig_LMC_AGB_Lensed}). 
      The inset shows a zoomed-in portion of the lensed LF around the knee of the lensed TRGB with dark blue dots showing the LF for $h=0.68$.  The two vertical dashed lines mark the position of the knee for each cosmology.}
         \label{Fig_LF_Lensed}
\end{figure}

The transient events identified by \cite{Fudamoto2024} provide a first indication of the LF of the Dragon's stars.  Figure~\ref{Fig_LF_Lensed} compares the result to Sect.~\ref{Sec_CMD}'s predictions. 
The two LFs were computed independently with normalization by the LMC's stellar surface density. The agreement in normalization is an interesting coincidence, but the agreement in slope suggests that the luminous LMC population is a good representation of the luminous Dragon population.
Different values of $h$ would move the predicted LF horizontally in the plot.  For  $h=0.68$,  stars at $z=0.725$ would be $8.8\%$ farther away (in luminosity distance) and be 0.18~mag fainter. For a power-law LF, such a shift could not be distinguished from a different stellar surface density (a vertical shift), but in fact the LMC LF has a ``knee'' at the TRGB jump, below which the number of stars dramatically increases. The knee occurs at $M_{J} \approx 39$ before magnification and $M_{J} \approx 30.5$ as observed. The observed magnitude of this knee depends on the value of $H_0$ as shown in the Figure~\ref{Fig_LF_Lensed} inset and is independent of the surface density of Dragon stars. 
Deep observations will detect a large number of Dragon stars and allow measuring the LF with great precision, thereby distinguishing between the two values of $h$.

\section{Discussion}\label{Sec_Discussion}
Different values of $H_0$ produce only modest differences in the observed LF, but Sect.~\ref{Sec_LF} has shown how the sharp TRGB feature survives after being magnified. However, measuring its position requires JWST images reaching 30.6~mag in F200W\null. The JAGB feature corresponds to brighter stars, but the feature is broader and harder to recognize in the LF\null. However, a careful selection of carbon AGB stars based on their color or other spectral features may allow using them as NESCs at this redshift.

Cepheids could be valuable NESCs because they are bright and can be detected at more modest magnification factors than TRGB or JAGB features. In addition, Cepheids' periodic nature may expose their identities when there are sufficient observations to construct their light curves. 
A challenge for identifying Cepheids will be the contaminating effect of microlenses, which will alter the Cepheid light curves. In most cases, these perturbations will be small because the timescale of variability of Cepheids (days to weeks) is much shorter than the typical timescale of microlensing (years) if we ignore the easily identifiable caustic-crossing events. In the rare occasions where a Cepheid is close to crossing a microcaustic, the change of magnification can take place in days, which can be identified thanks to the known dependency of magnification with time in microlensing, $\mu_{\rm micro}\propto |{t-t_0}|^{-0.5}$, where $t$ is time and $t_0$ is the crossing time.  

Unlike the JAGB or TRGB methods, a single Cepheid detection gives a distance, and the uncertainties decrease as $N^{-0.5}$ for $N$ stars. In contrast,  JAGB and TRGB stars inherently require statistics to detect the knee---the maximum steepness in their integral object counts---in the LF \citep{Freedman, Carleton}.
Therefore, to use JAGB and TRGB stars, the sample needs to go deep enough with sufficient statistics  to detect the knee. For lensed Cepheids, 
perhaps the most serious challenge is the unknown macromodel magnification, which is degenerate with the distance modulus that one is trying to measure. This degeneracy can be broken if multiple Cepheids are observed in close proximity, in which case the unknown magnifications can be constrained from the canonical $\mum=A/d$ law. At worst, Cepheids can be used to constrain $\mum$ at their positions, which in turn  improves the precision of the lens model and thus  the forward modeling of the observed LF\null.

\begin{figure} 
  \includegraphics[width=9.0cm]{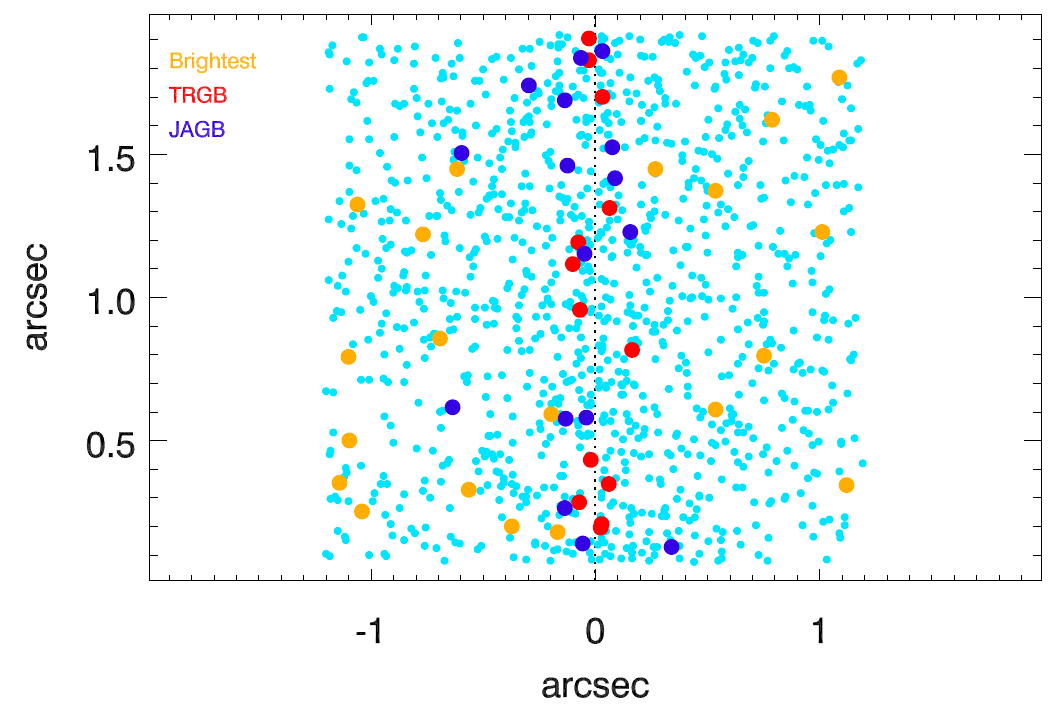}  
      \caption{Simulated spatial distribution of lensed events. The vertical dashed line represents the cluster CC\null.  The points represent all 1122 stars  above 30.6~mag in Figures~\ref{Fig_LMC_AGB_Lensed} and~\ref{Fig_LF_Lensed}. Red and dark blue dots represent detected stars in the TRGB and JAGB regions respectively. Orange dots represent the most luminous stars in the LMC ($M_{J}<34$ before lensing and $M_{J}<30.6$ after lensing). Cyan points represent other stars, mostly highly microlensed red giants intrinsically fainter than the TRGB\null. The $x$ location of each star was derived from a simple model where $\mum$ scales with $d$ as $\mum=60\arcsec/d$.  This gives $\mum=50$ at the maximum $d$ illustrated. The $y$ coordinate shown is a random value between 0\farcs1 and 1\farcs9, but this is a compressed scale.  The Dragon arc intersects the cluster CC at multiple locations, and the effective $y$ axis is $\approx$10\arcsec\ long.
      }
         \label{Fig_SpatialDistribution}
\end{figure}

An interesting point to consider is the spatial distribution of detected stars. Naturally the faintest stars require the largest magnification factors to be detected, while intrinsically brighter stars can be detected at more modest magnifications. The total magnification correlates with the macromodel magnification so we expect the faintest stars to concentrate  around the cluster CC as illustrated in Figure~\ref{Fig_SpatialDistribution}.
The TRGB stars are detected only near the cluster CC whereas stars in the center of the JAGB region are intrinsically brighter and can be detected farther from the CC\null. The most luminous stars hardly concentrate near the CC at all. Overall, the highest density of events is found at the CC, as expected for a relatively steep LF \citep{Diego2024b}. Given the similarity of the LMC LF to the observed one (Figure~\ref{Fig_LF_Lensed}), stars detected in  observations of the Dragon reaching magnitude 30.6 should trace the cluster CC with precision similar to that shown in Figure~\ref{Fig_SpatialDistribution}. More importantly, the faintest detections will also trace the smaller CCs of dark matter substructures  away from the main cluster CC\null. This will effectively map dark matter to very small scales  \citep{Diego2024b}.

Some of the brightest Dragon stars can appear as transients even without microlensing.  This is the case for Cepheids, which vary by a measurable amount within days or weeks. Detecting Cepheids in the Dragon offers a new avenue to constrain $H_0$ in a redshift range never tested before with this type of standard candle. As mentioned above, a challenge will be the unknown magnification, but this can be addressed statistically if sufficient Cepheids are observed. Alternatively, Cepheids can constrain the magnification at their locations, and these constraints can be applied iteratively to improve lens models and therefore distances based on other NESCs. The  details and limitations of such an iterative study are beyond the scope of this paper but will be addressed in future work.

\begin{figure} 
   \includegraphics[width=9.0cm]{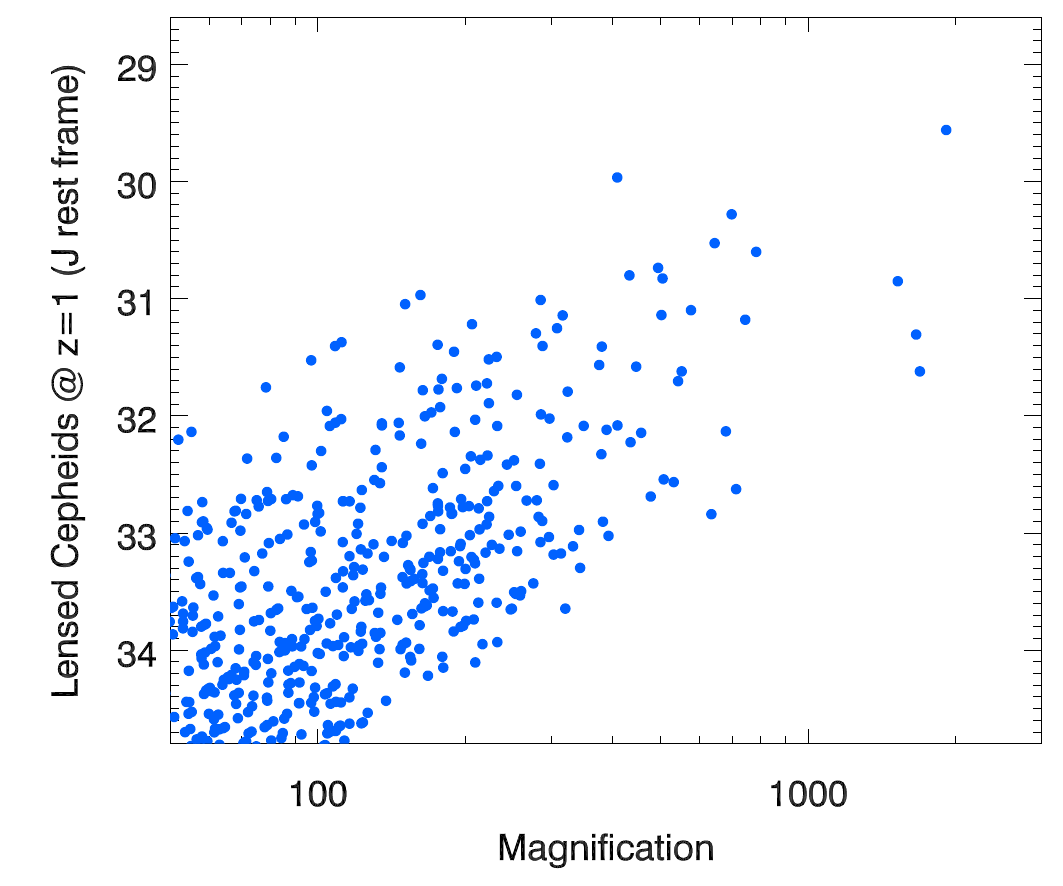}  
      \caption{Lensed Cepheids in the Spock arc at $z=1.0054$. The plot shows 10 magnification realizations of the 70 luminous Cepheid stars in the LMC \citep{Riess2019} but placed at the redshift of the Spock arc and magnified by the galaxy cluster plus microlenses. The $J$-band redshifts to 2.5~\micron, i.e., between JWST's F200W and F277W bands.}
         \label{Fig_Cepheids}
\end{figure}

If JWST can detect Cepheids in the Dragon, how much farther away can JWST detect them? The Spock arc, a $z=1.0054$ galaxy that is highly magnified by the galaxy cluster MACS0416, offers a case in point. Spock's luminosity distance is $\approx$50\%  larger than the Dragon's, and the increase in distance modulus is $\approx$0.9~mag.
To focus on the brightest Cepheids in the central region of the LMC, we adopted the  \cite{Riess2019} sample containing 70 luminous Cepheids. This sample should represent the small region of the Spock galaxy that is strongly magnified. From the $I$ and $H$ measurements \citep{Riess2019}, we constructed $J$-band magnitudes by a simple average and transformed them to AB\null. These stars were then placed at $z=1.0054$ and magnified by the same probabilities as in Figure~\ref{Fig_PDFmu}. The sample was simulated 10 times with sufficient separation in time (years) to make the magnitudes uncorrelated. The resulting apparent magnitudes are shown in Figure~\ref{Fig_Cepheids}.
From this figure, highly magnified long-period Cepheids in Spock would be accessible with regular medium-depth observations ($\approx$2~hour exposures in F200W or shorter with F150W2). Magnifications larger than the few thousand shown in this figure are still possible, so in principle Cepheids could be observed in Spock during a microlensing event with even shorter exposures. These events are often short-lived, with large magnification factors lasting a few weeks, but in some scenarios, large magnification factors can be maintained during several years. For example, a dark matter substructure or one of the many thousands of globular clusters in the ICM can produce long-term high magnifications. This type of rare scenario is suspected to be responsible for the long-lasting high-magnification of stars such as Godzilla \citep{DiegoGodzilla} and Mothra \citep{DiegoMothra}. The high magnification factors  in these two cases ($\mu>2000$) would make a bright Cepheid detectable for several years, allowing  measurement of its period (after subtracting microlensing distortions) and intrinsic luminosity. \\

Although challenging, the possibility of detecting Cepheids up to $z=1$ is an exciting, which may have been realized already as several transients in the Spock arc have been identified first  with HST \citep{Rodney2018,Kelly2022} and more recently with JWST \cite{Yan2023}. Additional stellar transients have been found behind the same cluster in a different (and much larger) spiral galaxy Warhol at $z=0.9396$ \citep{Chen2019,Kaurov2019,Yan2023}, where the Cepheids  in Figure~\ref{Fig_Cepheids} would be $\approx$0.2~mag brighter. 
Some of these transients may have been due to Cepheid variability. Long-term monitoring of these arcs will determine whether some of these transients are Cepheids, in which case they should reappear at regular times. Microlensing distortions to the magnification will certainly contaminate the light curves, but these are either subtle, with long-term variations extending years, or sudden if the Cepheid star crosses a microcaustic, with magnification which can be well modeled (and recognized) by the canonical law $\mu\propto {|t-t_o|}^{-0.5}$.

\section{Conclusions}\label{Sec_Conclusions}

We have simulated how the brightest stars in the LMC would appear if they are placed in the Dragon galaxy at redshift $z=0.725$ and amplified by the galaxy cluster A370 and microlenses in  the cluster's ICM\null. The large magnification factors near the cluster CC combined with microlensing  amplify the flux enough to make $\ga$1100 stars  detectable in observations reaching 30.6~mag in F200W\null.  This depth is within JWST's reach, and Dragon stars would be detected as transients (microlensing events) in difference images taken $\approx$1~year apart. The large number of stars allows accurate tracing of the the cluster's CC\null. More importantly, the faintest stars, such as the RGBs, will concentrate  around the position of invisible DM substructures, which create their own small CC\null. This will allow mapping the DM distribution down to  scales of $10^4$\Msun. 

Even accounting for the random effects of lensing, the observed LF of Dragon stars is expected change slope at $\approx$30.5~mag. This feature is due to the large number of RGB stars that start to be detected at this magnitude, and its  observed magnitude will differ by 0.18~mag for $H_0 =68$ versus 74. Galaxies at lower redshifts would show the LF break at brighter magnitudes, for instance 30.1~mag at $z=0.6$. Cepheids should be also easily detectable at these magnitudes. Regular observations of the Dragon (or similar strongly lensed arcs) will allow  construction of Cepheid light curves and deriving  periods and luminosities. The degeneracy between macromodel magnification and distance modulus can be broken by combining several Cepheids from the same galaxy opening the door to using them as standard candles at high redshift.  

Cepheid stars may be detectable even up to $z\approx1$. Galaxies such as Warhol and Spock already provide candidate Cepheids at this redshift.

\begin{acknowledgements}
 J.M.D. acknowledges support from project PID2022-138896NB-C51 (MCIU/AEI/MINECO/FEDER, UE) Ministerio de Ciencia, Investigaci\'on y Universidades.  
 RAW acknowledges support from NASA JWST Interdisciplinary Scientist grants
 NAG5-12460, NNX14AN10G and 80NSSC18K0200 from GSFC.
\end{acknowledgements}

\bibliographystyle{aa} 
\bibliography{MyBiblio} 

\begin{thebibliography}{28}
\expandafter\ifx\csname natexlab\endcsname\relax\def\natexlab#1{#1}\fi

\bibitem[{{Anand} {et~al.}(2024){Anand}, {Riess}, {Yuan}, {Beaton}, {Casertano}, {Li}, {Makarov}, {Makarova}, {Tully}, {Anderson}, {Breuval}, {Dolphin}, {Karachentsev}, {Macri}, \& {Scolnic}}]{Anand2024}
{Anand}, G.~S., {Riess}, A.~G., {Yuan}, W., {et~al.} 2024, ApJ, 966, 89

\bibitem[{{Chen} {et~al.}(2019){Chen}, {Kelly}, {Diego}, {Oguri}, {Williams}, {Zitrin}, {Treu}, {Smith}, {Broadhurst}, {Kaiser}, {Foley}, {Filippenko}, {Salo}, {Hjorth}, \& {Selsing}}]{Chen2019}
{Chen}, W., {Kelly}, P.~L., {Diego}, J.~M., {et~al.} 2019, \apj, 881, 8

\bibitem[{{Cioni} {et~al.}(2006){Cioni}, {Girardi}, {Marigo}, \& {Habing}}]{Cioni2006}
{Cioni}, M. R.~L., {Girardi}, L., {Marigo}, P., \& {Habing}, H.~J. 2006, A\&A, 448, 77

\bibitem[{{Diego} {et~al.}(2024{\natexlab{a}}){Diego}, {Kei Li}, {Amruth}, {Meena}, {Broadhurst}, {Kelly}, {Filippenko}, {Williams}, {Zitrin}, {Harris}, {Reina-Campos}, {Giocoli}, {Dai}, {Struble}, {Treu}, {Fudamoto}, {Gilman}, {Koekemoer}, {Lim}, {Palencia}, {Sun}, \& {Windhorst}}]{Diego2024a}
{Diego}, J.~M., {Kei Li}, S., {Amruth}, A., {et~al.} 2024{\natexlab{a}}, A\&A, 689, A167

\bibitem[{{Diego} {et~al.}(2024{\natexlab{b}}){Diego}, {Li}, {Amruth}, {Meena}, {Broadhurst}, {Kelly}, {Filippenko}, {Williams}, {Zitrin}, {Harris}, {Reina-Campos}, {Giocoli}, {Dai}, {Struble}, {Treu}, {Fudamoto}, {Gilman}, {Koekemoer}, {Lim}, {Palencia}, {Sun}, \& {Windhorst}}]{Diego2024b}
{Diego}, J.~M., {Li}, S.~K., {Amruth}, A., {et~al.} 2024{\natexlab{b}}, arXiv e-prints, arXiv:2404.08033

\bibitem[{{Diego} {et~al.}(2023{\natexlab{a}}){Diego}, {Meena}, {Adams}, {Broadhurst}, {Dai}, {Coe}, {Frye}, {Kelly}, {Koekemoer}, {Pascale}, {Willner}, {Zackrisson}, {Zitrin}, {Windhorst}, {Cohen}, {Jansen}, {Summers}, {Tompkins}, {Conselice}, {Driver}, {Yan}, {Grogin}, {Marshall}, {Pirzkal}, {Robotham}, {Ryan}, {Willmer}, {Bradley}, {Caminha}, {Caputi}, {Carleton}, \& {Kamieneski}}]{Diego2023Gordo}
{Diego}, J.~M., {Meena}, A.~K., {Adams}, N.~J., {et~al.} 2023{\natexlab{a}}, \aap, 672, A3

\bibitem[{{Diego} {et~al.}(2022){Diego}, {Pascale}, {Kavanagh}, {Kelly}, {Dai}, {Frye}, \& {Broadhurst}}]{DiegoGodzilla}
{Diego}, J.~M., {Pascale}, M., {Kavanagh}, B.~J., {et~al.} 2022, \aap, 665, A134

\bibitem[{{Diego} {et~al.}(2023{\natexlab{b}}){Diego}, {Sun}, {Yan}, {Furtak}, {Zackrisson}, {Dai}, {Kelly}, {Nonino}, {Adams}, {Meena}, {Willner}, {Zitrin}, {Cohen}, {D'Silva}, {Jansen}, {Summers}, {Windhorst}, {Coe}, {Conselice}, {Driver}, {Frye}, {Grogin}, {Koekemoer}, {Marshall}, {Pirzkal}, {Robotham}, {Rutkowski}, {Ryan}, {Tompkins}, {Willmer}, \& {Bhatawdekar}}]{DiegoMothra}
{Diego}, J.~M., {Sun}, B., {Yan}, H., {et~al.} 2023{\natexlab{b}}, \aap, 679, A31

\bibitem[{{Freedman} {et~al.}(2024){Freedman}, {Madore}, {Jang}, {Hoyt}, {Lee}, \& {Owens}}]{Freedman2024}
{Freedman}, W.~L., {Madore}, B.~F., {Jang}, I.~S., {et~al.} 2024, arXiv e-prints, arXiv:2408.06153

\bibitem[{{Fudamoto} {et~al.}(2024){Fudamoto}, {Sun}, {Diego}, {Dai}, {Oguri}, {Zitrin}, {Zackrisson}, {Jauzac}, {Lagattuta}, {Egami}, {Iani}, {Windhorst}, {Abe}, {Bauer}, {Bian}, {Bhatawdekar}, {Broadhurst}, {Cai}, {Chen}, {Chen}, {Cohen}, {Conselice}, {Espada}, {Foo}, {Frye}, {Fujimoto}, {Furtak}, {Golubchik}, {Hsiao}, {Jolly}, {Kawai}, {Kelly}, {Koekemoer}, {Kohno}, {Kokorev}, {Li}, {Li}, {Lin}, {Magdis}, {Meena}, {Nabizadeh}, {Richard}, {Steinhardt}, {Wu}, {Zhu}, \& {Zou}}]{Fudamoto2024}
{Fudamoto}, Y., {Sun}, F., {Diego}, J.~M., {et~al.} 2024, arXiv e-prints, arXiv:2404.08045

\bibitem[{{Furtak} {et~al.}(2024){Furtak}, {Meena}, {Zackrisson}, {Zitrin}, {Brammer}, {Coe}, {Diego}, {Eldridge}, {Jim{\'e}nez-Teja}, {Kokorev}, {Ricotti}, {Welch}, {Windhorst}, {Abdurro'uf}, {Andrade-Santos}, {Bhatawdekar}, {Bradley}, {Broadhurst}, {Chen}, {Conselice}, {Dayal}, {Frye}, {Fujimoto}, {Hsiao}, {Kelly}, {Mahler}, {Mandelker}, {Norman}, {Oguri}, {Pirzkal}, {Postman}, {Ravindranath}, {Vanzella}, \& {Wilkins}}]{Furtak2024}
{Furtak}, L.~J., {Meena}, A.~K., {Zackrisson}, E., {et~al.} 2024, \mnras, 527, L7

\bibitem[{{Kaurov} {et~al.}(2019){Kaurov}, {Dai}, {Venumadhav}, {Miralda-Escud{\'e}}, \& {Frye}}]{Kaurov2019}
{Kaurov}, A.~A., {Dai}, L., {Venumadhav}, T., {Miralda-Escud{\'e}}, J., \& {Frye}, B. 2019, \apj, 880, 58

\bibitem[{{Kelly} {et~al.}(2022){Kelly}, {Chen}, {Alfred}, {Broadhurst}, {Diego}, {Emami}, {Filippenko}, {Keen}, {Kei Li}, {Lim}, {Meena}, {Oguri}, {Scarlata}, {Treu}, {Williams}, {Williams}, {Zhou}, {Zitrin}, {Foley}, {Jha}, {Kaiser}, {Mehta}, {Rieck}, {Salo}, {Smith}, \& {Weisz}}]{Kelly2022}
{Kelly}, P.~L., {Chen}, W., {Alfred}, A., {et~al.} 2022, arXiv e-prints, arXiv:2211.02670

\bibitem[{{Kelly} {et~al.}(2018){Kelly}, {Diego}, {Rodney}, {Kaiser}, {Broadhurst}, {Zitrin}, {Treu}, {P{\'e}rez-Gonz{\'a}lez}, {Morishita}, {Jauzac}, {Selsing}, {Oguri}, {Pueyo}, {Ross}, {Filippenko}, {Smith}, {Hjorth}, {Cenko}, {Wang}, {Howell}, {Richard}, {Frye}, {Jha}, {Foley}, {Norman}, {Bradac}, {Zheng}, {Brammer}, {Benito}, {Cava}, {Christensen}, {de Mink}, {Graur}, {Grillo}, {Kawamata}, {Kneib}, {Matheson}, {McCully}, {Nonino}, {P{\'e}rez-Fournon}, {Riess}, {Rosati}, {Schmidt}, {Sharon}, \& {Weiner}}]{Kelly2018}
{Kelly}, P.~L., {Diego}, J.~M., {Rodney}, S., {et~al.} 2018, Nature Astronomy, 2, 334

\bibitem[{{Madore} \& {Freedman}(2020)}]{Madore2020}
{Madore}, B.~F. \& {Freedman}, W.~L. 2020, ApJ, 899, 66

\bibitem[{{Ma{\'\i}z Apell{\'a}niz}(2007)}]{Maiz2007}
{Ma{\'\i}z Apell{\'a}niz}, J. 2007, in Astronomical Society of the Pacific Conference Series, Vol. 364, The Future of Photometric, Spectrophotometric and Polarimetric Standardization, ed. C.~{Sterken}, 227

\bibitem[{{Newman} {et~al.}(2024){Newman}, {McQuinn}, {Skillman}, {Boyer}, {Cohen}, {Dolphin}, \& {Telford}}]{Newman2024}
{Newman}, M. J.~B., {McQuinn}, K. B.~W., {Skillman}, E.~D., {et~al.} 2024, arXiv e-prints, arXiv:2406.03532

\bibitem[{{Palencia} {et~al.}(2024{\natexlab{a}}){Palencia}, {Diego}, {Kavanagh}, \& {Mart{\'\i}nez-Arrizabalaga}}]{Palencia2023}
{Palencia}, J.~M., {Diego}, J.~M., {Kavanagh}, B.~J., \& {Mart{\'\i}nez-Arrizabalaga}, J. 2024{\natexlab{a}}, \aap, 687, A81

\bibitem[{{Palencia} {et~al.}(2024{\natexlab{b}}){Palencia}, {Diego}, {Kavanagh}, \& {Mart{\'\i}nez-Arrizabalaga}}]{Palencia2024}
{Palencia}, J.~M., {Diego}, J.~M., {Kavanagh}, B.~J., \& {Mart{\'\i}nez-Arrizabalaga}, J. 2024{\natexlab{b}}, A\&A, 687, A81

\bibitem[{{Planck Collaboration} {et~al.}(2020){Planck Collaboration}, {Aghanim}, {Akrami}, {Ashdown}, {Aumont}, {Baccigalupi}, {Ballardini}, {Banday}, {Barreiro}, {Bartolo}, {Basak}, {Battye}, {Benabed}, {Bernard}, {Bersanelli}, {Bielewicz}, {Bock}, {Bond}, {Borrill}, {Bouchet}, {Boulanger}, {Bucher}, {Burigana}, {Butler}, {Calabrese}, {Cardoso}, {Carron}, {Challinor}, {Chiang}, {Chluba}, {Colombo}, {Combet}, {Contreras}, {Crill}, {Cuttaia}, {de Bernardis}, {de Zotti}, {Delabrouille}, {Delouis}, {Di Valentino}, {Diego}, {Dor{\'e}}, {Douspis}, {Ducout}, {Dupac}, {Dusini}, {Efstathiou}, {Elsner}, {En{\ss}lin}, {Eriksen}, {Fantaye}, {Farhang}, {Fergusson}, {Fernandez-Cobos}, {Finelli}, {Forastieri}, {Frailis}, {Fraisse}, {Franceschi}, {Frolov}, {Galeotta}, {Galli}, {Ganga}, {G{\'e}nova-Santos}, {Gerbino}, {Ghosh}, {Gonz{\'a}lez-Nuevo}, {G{\'o}rski}, {Gratton}, {Gruppuso}, {Gudmundsson}, {Hamann}, {Handley}, {Hansen}, {Herranz}, {Hildebrandt}, {Hivon}, {Huang}, {Jaffe}, {Jones}, {Karakci}, {Keih{\"a}nen},
  {Keskitalo}, {Kiiveri}, {Kim}, {Kisner}, {Knox}, {Krachmalnicoff}, {Kunz}, {Kurki-Suonio}, {Lagache}, {Lamarre}, {Lasenby}, {Lattanzi}, {Lawrence}, {Le Jeune}, {Lemos}, {Lesgourgues}, {Levrier}, {Lewis}, {Liguori}, {Lilje}, {Lilley}, {Lindholm}, {L{\'o}pez-Caniego}, {Lubin}, {Ma}, {Mac{\'\i}as-P{\'e}rez}, {Maggio}, {Maino}, {Mandolesi}, {Mangilli}, {Marcos-Caballero}, {Maris}, {Martin}, {Martinelli}, {Mart{\'\i}nez-Gonz{\'a}lez}, {Matarrese}, {Mauri}, {McEwen}, {Meinhold}, {Melchiorri}, {Mennella}, {Migliaccio}, {Millea}, {Mitra}, {Miville-Desch{\^e}nes}, {Molinari}, {Montier}, {Morgante}, {Moss}, {Natoli}, {N{\o}rgaard-Nielsen}, {Pagano}, {Paoletti}, {Partridge}, {Patanchon}, {Peiris}, {Perrotta}, {Pettorino}, {Piacentini}, {Polastri}, {Polenta}, {Puget}, {Rachen}, {Reinecke}, {Remazeilles}, {Renzi}, {Rocha}, {Rosset}, {Roudier}, {Rubi{\~n}o-Mart{\'\i}n}, {Ruiz-Granados}, {Salvati}, {Sandri}, {Savelainen}, {Scott}, {Shellard}, {Sirignano}, {Sirri}, {Spencer}, {Sunyaev}, {Suur-Uski}, {Tauber}, {Tavagnacco},
  {Tenti}, {Toffolatti}, {Tomasi}, {Trombetti}, {Valenziano}, {Valiviita}, {Van Tent}, {Vibert}, {Vielva}, {Villa}, {Vittorio}, {Wandelt}, {Wehus}, {White}, {White}, {Zacchei}, \& {Zonca}}]{Planck2020}
{Planck Collaboration}, {Aghanim}, N., {Akrami}, Y., {et~al.} 2020, \aap, 641, A6

\bibitem[{{Riess} {et~al.}(2019){Riess}, {Casertano}, {Yuan}, {Macri}, \& {Scolnic}}]{Riess2019}
{Riess}, A.~G., {Casertano}, S., {Yuan}, W., {Macri}, L.~M., \& {Scolnic}, D. 2019, \apj, 876, 85

\bibitem[{{Ripoche} {et~al.}(2020){Ripoche}, {Heyl}, {Parada}, \& {Richer}}]{Ripoche2020}
{Ripoche}, P., {Heyl}, J., {Parada}, J., \& {Richer}, H. 2020, MNRAS, 495, 2858

\bibitem[{{Rodney} {et~al.}(2018){Rodney}, {Balestra}, {Bradac}, {Brammer}, {Broadhurst}, {Caminha}, {Chiriv{\i}}, {Diego}, {Filippenko}, {Foley}, {Graur}, {Grillo}, {Hemmati}, {Hjorth}, {Hoag}, {Jauzac}, {Jha}, {Kawamata}, {Kelly}, {McCully}, {Mobasher}, {Molino}, {Oguri}, {Richard}, {Riess}, {Rosati}, {Schmidt}, {Selsing}, {Sharon}, {Strolger}, {Suyu}, {Treu}, {Weiner}, {Williams}, \& {Zitrin}}]{Rodney2018}
{Rodney}, S.~A., {Balestra}, I., {Bradac}, M., {et~al.} 2018, Nature Astronomy, 2, 324

\bibitem[{{Schneider} {et~al.}(1992){Schneider}, {Ehlers}, \& {Falco}}]{Schneider1992}
{Schneider}, P., {Ehlers}, J., \& {Falco}, E.~E. 1992, {Gravitational Lenses, XIV} ({Springer-Verlag})

\bibitem[{{Storm} {et~al.}(2011){Storm}, {Gieren}, {Fouqu{\'e}}, {Barnes}, {Pietrzy{\'n}ski}, {Nardetto}, {Weber}, {Granzer}, \& {Strassmeier}}]{Storm2011}
{Storm}, J., {Gieren}, W., {Fouqu{\'e}}, P., {et~al.} 2011, A\&A, 534, A94

\bibitem[{{Weinberg} \& {Nikolaev}(2001)}]{Weinberg2001}
{Weinberg}, M.~D. \& {Nikolaev}, S. 2001, ApJ, 548, 712

\bibitem[{{Welch} {et~al.}(2022){Welch}, {Coe}, {Diego}, {Zitrin}, {Zackrisson}, {Dimauro}, {Jim{\'e}nez-Teja}, {Kelly}, {Mahler}, {Oguri}, {Timmes}, {Windhorst}, {Florian}, {de Mink}, {Avila}, {Anderson}, {Bradley}, {Sharon}, {Vikaeus}, {McCandliss}, {Brada{\v{c}}}, {Rigby}, {Frye}, {Toft}, {Strait}, {Trenti}, {Sharma}, {Andrade-Santos}, \& {Broadhurst}}]{Welch2022}
{Welch}, B., {Coe}, D., {Diego}, J.~M., {et~al.} 2022, \nat, 603, 815

\bibitem[{{Yan} {et~al.}(2023){Yan}, {Ma}, {Sun}, {Wang}, {Kelly}, {Diego}, {Cohen}, {Windhorst}, {Jansen}, {Grogin}, {Beacom}, {Conselice}, {Driver}, {Frye}, {Coe}, {Marshall}, {Koekemoer}, {Willmer}, {Robotham}, {D'Silva}, {Summers}, {Nonino}, {Pirzkal}, {Ryan}, {Ortiz}, {Tompkins}, {Bhatawdekar}, {Cheng}, {Zitrin}, \& {Willner}}]{Yan2023}
{Yan}, H., {Ma}, Z., {Sun}, B., {et~al.} 2023, \apjs, 269, 43

\end{thebibliography}


\end{document}